\documentclass[a4paper,11pt,twoside]{article}
\usepackage{fancyhdr,latexsym,amsmath,amsfonts,amssymb,amsbsy,amsthm,url}%
\usepackage[hscale=0.7,centering]{geometry}
\usepackage{graphicx,epsfig, xy}
\allowdisplaybreaks

\theoremstyle{definition}

\theoremstyle{remark}

\begin{document}
\title
{Modelling COVID-19 - I\\
A dynamic SIR(D) with application to Indian data}
\author{
\parbox[t]{0.60\textwidth}{Madhuchhanda Bhattacharjee\footnote{Research supported by a SERB-MATRICS grant for COVID-19.}\\
School of Mathematics and Statistics\\ University of Hyderabad\\ Gachibowli\\ Hyderabad 500046\\ India\\
email: mbsm@uohyd.ernet.in}
\hspace{.05\textwidth}
\parbox[t]{0.45\textwidth}{Arup Bose\footnote{Research supported by J.C. Bose National Fellowship, Dept.~of Science and Tech., Govt.~of India.}\\
Stat-Math Unit, Kolkata\\
Indian Statistical Institute\\
203 B.T. Road\\
Kolkata 700108\\ India\\
email: bosearu@gmail.com}
}
\date{\today}

\maketitle
\vspace{-40pt}
\begin{abstract} \noindent
{\footnotesize We propose an epidemiological model using an adaptive dynamic three compartment (with four states) SIR(D) model. Our approach is similar to non-parametric curve fitting in spirit and automatically adapts to key external factors, such as interventions, while retaining the parsimonious nature of the standard SIR(D) model.

Initial dynamic temporal estimates of the model parameters are obtained by minimising the aggregate residual sum of squares across the number of infections, recoveries, and fatalities, over a chosen lag period. Then a geometric smoother is applied to obtain the final time series of estimates. These estimates are used to obtain dynamic temporal robust estimates of the key feature of this pandemic, namely the ``reproduction number".

We illustrate our method on the Indian  COVID-19 data for the period March 14--August 31, 2020.
The time series data plots of the 36 states and union territories shows a clear presence of inter-regional variation in the prognosis of the epidemic. This is also bourne out by the estimates of the underlying parameters, including the reproduction numbers for the 36 regions.
Due to this, an SIR(D) model, dynamic or otherwise,  on the national aggregate data is not suited for robust local predictions.

The time series of estimates of the model enables us to carry out daily, weekly and also long term predictions, including construction of predictive bands. We obtain an excellent agreement between the actual data and the model predicted data at the regional level.

Our estimates of the current reproduction number turn out to be more than 2 in three regions (Andhra Pradesh, Maharashtra and Uttar Pradesh) and between 1.5 and 2 in 13 regions. Each of these regions have experienced an individual trajectory, which typically involves initial phase of shock(s) followed by a relatively steady lower level of the reproduction number.
}
\end{abstract}
\vskip-10pt
{\footnotesize \noindent   {\bf Keywords.} COVID-19, dynamic SIR(D) model, differential equation, least squares, estimation, prediction, reproduction number, curve-fitting, geometric smoothing, robust estimator,  \texttt{R}-code. }

{\footnotesize \noindent {\bf AMS 2000 Subject Classification.}
Primary 62P10, 
Secondary 92D30} 
\normalsize

\section{Introduction}\label{section:intro}
The basic deterministic SIR(D) (susceptible-infected-removed) model  for the evolution of an epidemic over time was proposed by Kermack and McKendrick in 1927 and has been widely used and accepted. It is a three compartment model that focuses on the \textit{effective contact rate} $\beta$, the \textit{recovery rate} $\nu$, and the \textit{fatality rate} $\mu$,  which are all assumed to remain constant throughout the epidemic.
While this model has been very fruitful, there have been numerous variations of it, spurred by a need to accommodate factors specific to a given epidemic.

Like all models, SIR(D) envisages an ideal situation and hence some of the primary assumptions governing the spread would always be violated in any real-life epidemic, and COVID-19 is no exception. For instance, there have been periodic local and national interventions as well as different levels of compliance to safety precautions. Many variations of the SIR(D) have been proposed and used in the context of COVID-19. See for example, Kucharski et al.~(2020), Liu et al.~(2020) and Sun et al.~(2020).

In particular, for India, there is clear evidence that the nature and extent of spread of COVID-19 has been varied widely across its 36 states and union territories (henceforth  collectively referred as ``regions''). This is evident from the various time series plots that we shall exihibit later. See also, for example, the recent article of Ranjan (2020). As a consequence, any modelling and analysis must be done at least region-wise, which is what we attempt in this article.

It is also important to investigate the spatio-temporal dependence in COVID-19 data. Cliff and Ord (1981) showed how taking spatial dependence into consideration can be useful in the statistical analysis of epidemics. One may be able to ``borrow strength'' from connected (not necessarily geographically) regions for an improved analysis. This significantly more challenging issue is not addressed in the present article.

We stick to the basic tenets of the time tested SIR(D) model. However, as mentioned above, there are underlying movers that disturb the ideal SIR(D) world. For India,  primary among these are{\textemdash}various lockdowns and unlocks, migration of people at different levels and, level of compliance with the guidelines. Further, there are issues with the collected data{\textemdash}under-reporting; delayed reporting; change in the methodology of reporting;  evolution of medical/clinical methods for case identification over time, including method of sample collection and laboratory testing procedure. There have been attempts by researchers to incorporate at least some of these factors into an SIR(D) model. See for example Ansumali and Meher (2020) and Kotwal et al.~(2020).

However, we do not try to delineate the effects of these external factors. We take a parsimonious approach and do not add further controlling parameters. We continue with the SIR(D) model but let it evolve adaptively with the temporal data. It is possible to do so in this case since, unlike other epidemics, for COVID-19,  in addition to number of infections, time series data is also available on number of recoveries and fatalities. In a broad sense our approach is akin to non-parametric curve fitting.

In Section \ref{sec:background}, we give a brief background on the  history and nature of spread of COVID-19 in India. In Section \ref{sec:SIR(D)} we first introduce the concept of the reproduction number. Then we explain briefly the basic SIR(D) model and the formula for the reproduction number for this model in terms of the model parameters. We also present our extension of this model, inspired by non-parametric curve fitting techniques, and present methods of estimating the relevant parameters. In Section \ref{sec:analysis} we provide a detailed description of the findings based on our analysis of the region-wise Indian data. Section \ref{sec:conclusions} concludes with discussions and some observations for future work.

\section{Material: COVID-19 in India}\label{sec:background}
\begin{figure}[h]
\begin{center}\includegraphics[height=65mm, width=110mm]
{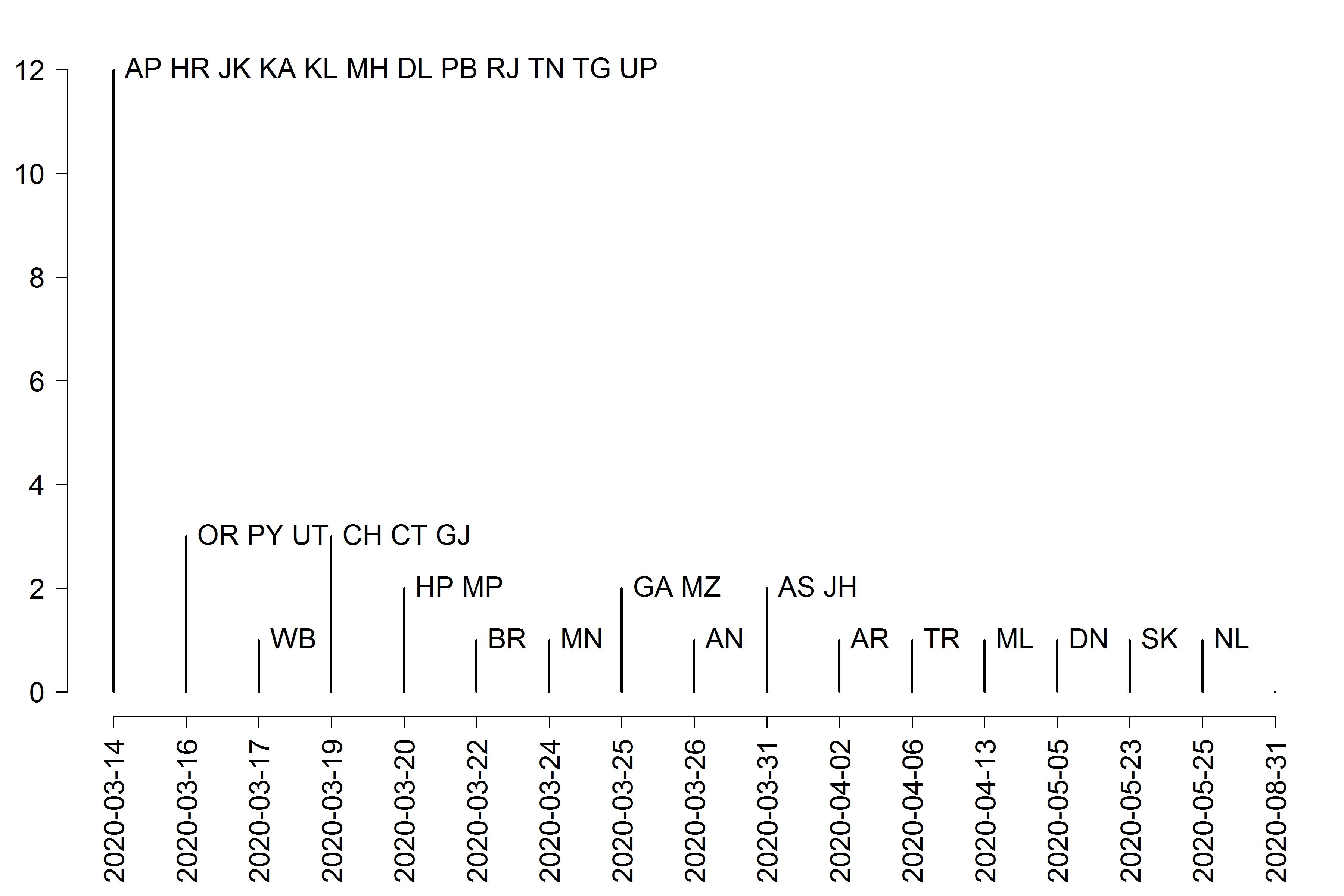}
\end{center}
\vskip-10pt
\small\caption{\small Regional pattern of ``initial'' appearance of COVID-19.}
\label{fig:First_case_date_for_states.pdf}
\end{figure}

\begin{figure}[h]
\vskip-10pt
\begin{center}\includegraphics[height=65mm, width=100mm]{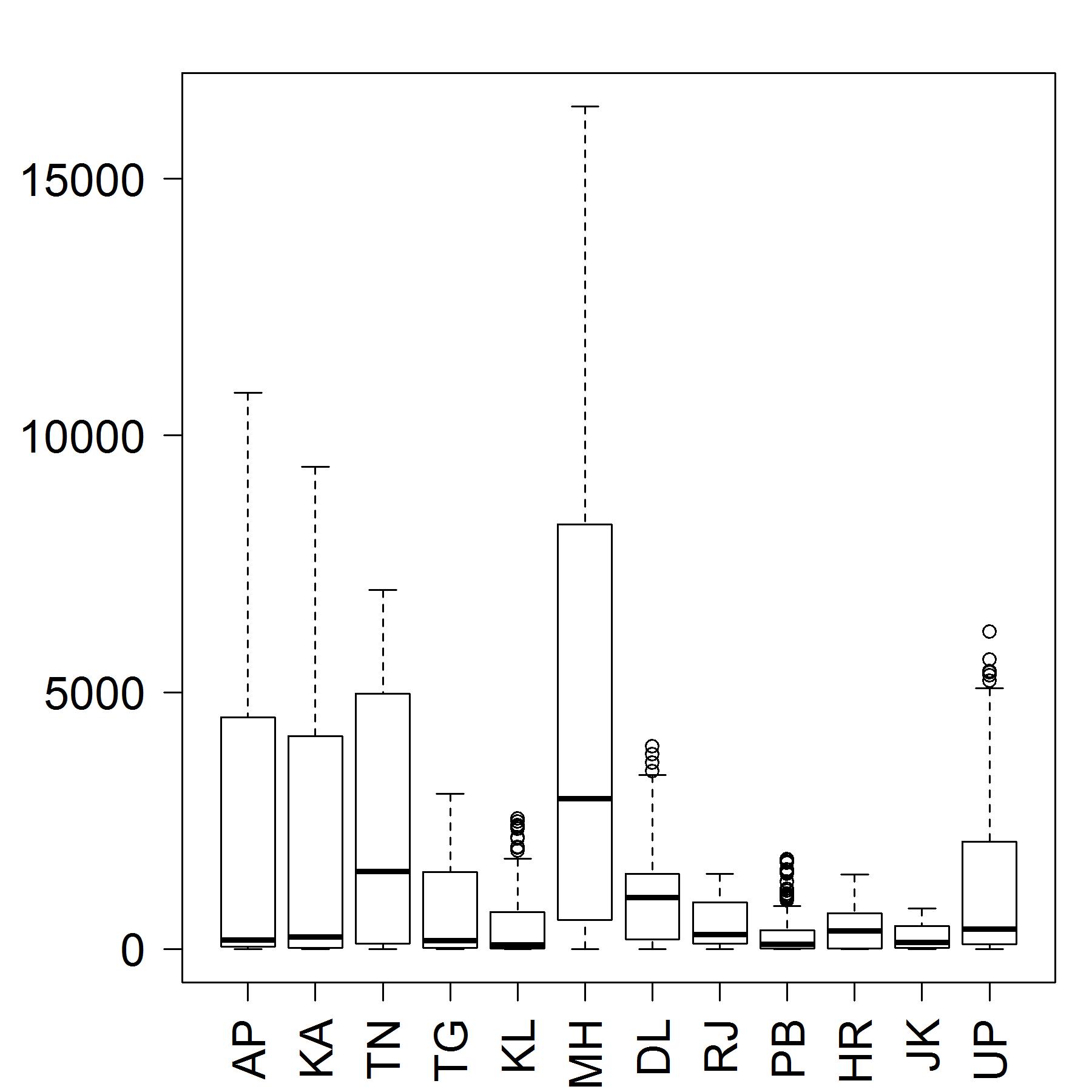}
\end{center}
\vskip-10pt
\caption{\small Distribution of daily infections, 12 first-hit regions, March 14--August 31, 2020}
\label{fig:First_12_state_daily_case_distn.pdf}
\vskip3pt
\end{figure}

\begin{figure}[h]
\begin{center}\includegraphics[height=100mm, width=100mm]{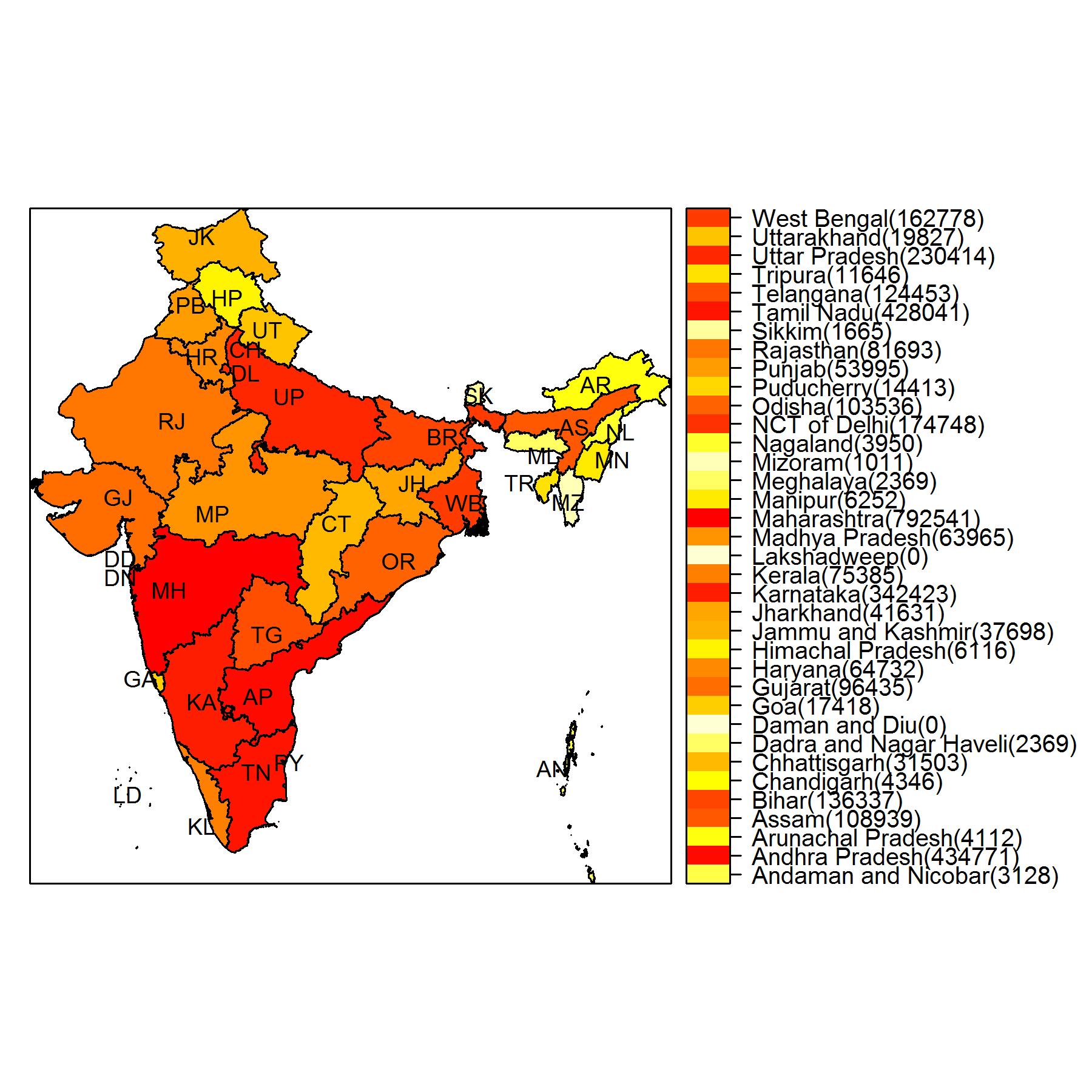}
\end{center}
\vskip-65pt
 \caption{\small Regional cumulative infection counts till August 31, 2020}
\label{fig:India_state_csum.jpg}
\end{figure}

We have used the daily data on number of confirmed, recovered and deceased cases from the 36 states and union territories (in short, regions) of India, for analyses and illustrations of our proposed model. Our primary data source is the volunteer-driven, crowd-sourced database \url{https://api.covid19india.org} for COVID-19 statistics and patient tracing in India. We have used the same abbreviated initials for the names of the regions as in this website. The full names are given in Table \ref{table:code}.\vskip-10pt
\small
\begin{table}[ht!]
\begin{center}
\caption{\small Names and codes of regions of India}\label{table:code}
\vskip 3pt
{\footnotesize
\begin{tabular}{|c|l|c|l|c|l|}
\hline
Code& Region         & Code & Region                  & Code & Region\\ \hline
AP	& Andhra Pradesh & AR	  & Arunachal Pradesh       & AS	 & Assam \\
BR	& Bihar          & CH	  & Chandigarh              & CT	 & Chhattisgarh \\
DL	& Delhi          & DN	  & Dadra and Nagar Haveli, & GA	 & Goa \\
GJ  & Gujarat        &      & Daman and Diu           & HP   & Himachal Pradesh\\
HR	& Haryana        & JH	  & Jharkhand               & JK   & Jammu and Kashmir \\
KA	& Karnataka      & KL	  & Kerala                  & LA	 & Ladakh  \\
LD	& Lakshadweep    & MH	  & Maharashtra             & ML	 & Meghalaya \\
MN	& Manipur        & MP	  & Madhya Pradesh          & MZ	 & Mizoram  \\
NL	& Nagaland       & OR   & Odisha                  & PB	 & Punjab\\
PY	& Puducherry     & RJ   & Rajasthan               & SK	 & Sikkim \\
TG	& Telangana      & TN   & Tamil Nadu              & TR   & Tripura \\
UP	& Uttar Pradesh  & UT   & Uttarakhand             & WB	 & West Bengal\\
\hline
\end{tabular}}
\end{center}
\end{table}
\vskip-15pt
\normalsize
The first infection in India was reported on January 30, 2020. Prior to March 14, there were only sporadic and stray  appearances, of at most 100 confirmed cases across entire India, whose conclusive regional breakup is not clear. An overwhelming number of regions had none to a very low number of infections till then.  Figure \ref{fig:First_case_date_for_states.pdf} gives the dates of ``first appearance'' and the counts for different regions in the early period since March 14, when the number of infections saw a rapid increase. The extent and speed of spread across regions have been quite varying. See Figures \ref{fig:First_12_state_daily_case_distn.pdf} and \ref{fig:India_state_csum.jpg}. We thus restrict our analysis to daily regional data for the period March 14--August 31, 2020.

It is not the goal of this paper to identify and quantify the factors governing this regional heterogeneity. Nevertheless,
we present a short review of some of the more important factors that are relevant, borrowing from Wikipedia (2020), The Print (2020) 
and HT (2020a). 
\normalsize

\noindent \textbf{(i) Lock and unlock: salient features}. The Central and State Governments announced and implemented different containment measures such as complete and partial national lockdowns, local lockdowns, closure and re-opening of certain offices, businesses, schools, closure and partial reopening of domestic and international transport, prohibition on travel and outdoor activities. Table \ref{table:lockdowns} summarises the key intervention dates. We give some of the salient features of these interventions below.

\vskip3pt
\vskip-10pt
\small
\begin{table}[h]
\vskip-10pt
\begin{center}
\caption{\small Locks and unlocks}\label{table:lockdowns}
\vskip 3pt
{\footnotesize
\begin{tabular}{|c|c|c|}
\hline
Phase       & From       & To             \\
 \hline
Lock 1      &  March 25 & April 14        \\
Lock 2      &  April 15  & May 03         \\
Lock 3      &  May 04    & May 17         \\
Lock 4      &  May 18    & May 31         \\
\hline
Unlock   1  &  June 01   & June 30         \\
Unlock   2  &  July 01   & July 31        \\
Unlock   3  &  August 01 & August 31      \\
\hline
\end{tabular}}
\end{center}
\vskip-25pt
\end{table}
\normalsize

\noindent
\textit{Phase 1 (March 25 to April 14)}: On March 22, Indian Railways suspended passenger operations 
and on March 25, nearly all services and factories were suspended by a central Govt notification.


\noindent
\textit{Phase 2 (April 15 to May 03)}: On April 16, lockdown areas were classified as ``red zone'' which had infection hotspots, ``orange zone'' with moderate amount of infection, and ``green zone'' with no infections. On April 20, reopening was allowed for: agricultural businesses, aquaculture, plantations and farming supplies shops; public works programmes with social distancing; cargo transportation; banks and government centres distributing benefits. On April 25, opening of small retail shops was allowed with half the staff and social distancing. On April 29, inter-state movement of  stranded persons were allowed under suitable guidelines.

\noindent
\textit{Phase 3 (May 04 to May 17)}: The three zones were redefined: red (130 districts) with high COVID-19 cases and a high
doubling\footnote{Defined as the further number of days in which the current total number of infections will double.} rate;  orange (284 districts) with comparatively fewer cases than red;  green (319 districts)  without any cases in the past 21 days. The red zones remained under lockdown, only private and hired vehicles were allowed in  orange zones, and normal movement was permitted in green zones with buses limited to 50\%  capacity. The zone classification was monitored once a week.

\noindent
\textit{Phase 4 (May 18 to May 31)}: Regions were given a larger say in the demarcation of the three zones and their implementation.
Red zones were further divided into containment and buffer zones and local bodies were given the authority to demarcate these.

\noindent
\textit{Unlock 1.0 (June 01 to June 30)}:
Lockdown restrictions were imposed only in containment zones, while activities were permitted in other zones in a phased manner. From June 08, there was reopening of shopping malls, religious places, hotels and restaurants. Large gatherings were not permitted. Inter-state travel was permitted. Nightly curfew was imposed from 9 p.m.~to 5 a.m.~in all areas. State governments were allowed to impose suitable restrictions on all activities.

\noindent
\textit{Unlock 2.0 (July 01  to July 31)}: Lockdown measures  were imposed only in containment zones. In all other areas, most activities were permitted. Nightly curfew period was modified slightly.
State governments continued to be at liberty to impose suitable restrictions. State borders remained open. Inter- and intra-state travel was permitted. Limited international travel was permitted as part of the Vande Bharat Mission (see below). Shops were permitted to allow more than five persons at a time. Educational institutions, metros, recreational activities remained closed till July 31. Only essential activities were permitted in containment zones, while maintaining strict parameter control and ``intensive contact tracing, house-to-house surveillance, and other clinical interventions''.

\noindent
\textit{Unlock 3.0 (August 01 to August 31)}: Night curfews were removed and gymnasiums and yoga centres were permitted to reopen from  August 05. Educational institutions remained closed till August 31. All inter- and intra-state travel and transportation was permitted. Independence Day celebrations was permitted with social distancing. Maharashtra and Tamil Nadu imposed a lockdown for the whole month, while West Bengal imposed lockdowns twice a week and later made some adjustments.
\vskip5pt

In addition to the above, there have been other restrictions and guidelines that have affected the spread. For instance, the movement across Myanmar border along Mizoram and Manipur was restricted since March 09.

\noindent \textbf{(ii) Migration}: It appears that no definitive data is available on the (reverse) migration of workers who were left stranded in their places of work as businesses and other activities came to a standstill after the first lockdown was announced.

We gather from different news sources that soon after the national lockdown was announced, a first wave of migration begun around March 24, 2020 and lasted for about two weeks. Much of the inter-region migration appears to have taken place initially from Delhi, Gujarat, Haryana, Maharashtra and Punjab,  to  Assam, Bihar, Madhya Pradesh, Orissa, Rajasthan, Uttar Pradesh, and West Bengal.

Later the migration occurred between additional states; for example from Karnataka, Maharashtra, Tamil Nadu and West Bengal to Assam,  Manipur, Meghalaya and Tripura. This wave started around May 04, 2020 and again lasted for about two weeks.

After the first phase of unlock, many of the returnees began to go back to their places of work in different regions.

Estimates of the total number of people that moved during these migrations in the last six months vary widely,  but all of these run into several hundred thousands. There is evidence that the first group of returnees had an extremely low percentage of infected persons and hence had extremely low risk of spreading the virus. The later group of returnees had a much higher percentage of infected people and the risk of spread increased significantly. See for example Mohan and Amin (2020) for a report on a limited study in Rajasthan and The Wire (2020) for  news on spike in cases in the North-eastern states after May 04, 2020.

%

\noindent \textbf{(iii) International flights: Vande Bharat Mission}:  Regular international flights to and from India remain suspended since
March 22 till August 31, 2020 (see HT (2020b)). 
However, later about 2500 flights of foreign carriers were allowed to and from India, particularly for stranded foreign passengers.

The Vande Bharat mission of the Government of India was launched on May 07, 2020 to bring back stranded Indians from around the world. Special flights, both of Government and private airlines,  to and from designated cities, were flown for this purpose in five phases{\textemdash}May 07--15, 2020; May 17--June 10; June 11--July 02; July  15--31; August 01--31. A sixth phase is planned from September 01--October 24, 2020. Economic Times (2020) reports that as of August 20, over 11.2 lakh Indian nationals have returned from abroad on these flights. Stranded Indians have also returned by land from the neighbouring countries.

Again, the regional variation in the count of returning Indians have been significant. While quarantine protocols have been in place for incoming passengers, there have been laxity of varying degrees in adherence to the restrictions, thereby introducing another level of variation between the regions.
\vskip5pt

Additionally, there have been other measures such as, mandatory quarantine for inter-state travellers, home isolation for infected individuals and restriction of human movement around places of infections. Orders and advisories issued by local authorities were guided by local situations and hence were also varied. There have also been variations in the implementation of, and compliance to, all the different orders over time. Moreover, it is commonly accepted that there is non-uniformity in the manner the data has been collected as well as in its coverage and quality. Thus, the growth and spread has been quite varying across the 36 regions.
The models we shall propose shortly, are thus going to be implemented regionally.

\section{Model}\label{sec:SIR(D)}
Before we explain our model and procedures, we give a brief introduction to the basic SIR(D) model, including how the crucial concept of the reproduction number relates to the model parameters.

\subsection{Reproduction number and the SIR(D) model}

We may consider three essential parameters that drive \textit{any} epidemic: \vskip5pt

\noindent --the transmissibility parameter $\tau$. This is the probability of infection,  given contact between a susceptible
and an infected individual;

\noindent
--the average rate of contact between susceptible and infected individuals, say $\tilde c$;

\noindent --the duration of infectiousness, $d$. \vskip5pt

\noindent Simply put, the basic \textit{reproduction number}, usually denoted by $\mathcal{R}_0$ is the expected number of secondary cases produced by a single (typical) infection in a completely susceptible population. It is a dimensionless quantity and can be interpreted as
$$\mathcal{R}_0\propto \tau \tilde c d.$$
Its importance stems from the fact that it can be related to key quantities that dictate whether an epidemic progresses or dies  off.

The reproduction number has a direct interpretation in an SIR(D) model (\textit{Susceptible-Infected-Removed (Deceased)}). The ``removed'' group consists of those individuals who have recovered or have succumbed. This model for the propagation of an epidemic was originally introduced by Kermack and McKendrick in 1927, and along with several variations, have been reprinted as Kermack and McKendrick (1991a, b, c) due to their immense importance and use. Here is a brief introduction to this model.
\vskip5pt

\noindent \textbf{Assumption I}. \vskip5pt

\noindent
a. The epidemic propagates in a closed, completely susceptible, well mixed  population of a constant size. There is no demographic change (no net births/deaths) during the course of the propagation.

\noindent
b. The propagation happens at constant rates of infection, recovery and fatalities.  \vskip5pt


Let $N$ be the size of the population. At any time point $t$, let
\begin{eqnarray*} S(t)&=& \text{cumulative number of susceptible individuals}\\
I(t)&=& \text{cumulative number of infections}\\
R(t)&=& \text{cumulative number of recoveries}\\
D(t)&=& \text{cumulative number of fatalities}.
\end{eqnarray*}
Note that Assumption I(a) implies that
\begin{equation*}\label{constantpop}N=S(t)+I(t)+R(t)+D(t)\ \ \text{for all} \ \ t.
\end{equation*}
Often these counts are expressed as proportions and then we can assume that $N$ is normalized to $1$.

The SIR(D) model governing the propagation (over continuous time) of the epidemic honouring Assumption I is given by a simultaneous differential equation:
\begin{eqnarray}
dS(t)/dt &=& -\beta (S(t)/N) I(t)\label{eqns}\\
dI(t)/dt &=&\beta (S(t)/N) I(t) - \nu I(t)-\mu I(t)\label{eqni}\\
dR(t)/dt &=& \nu I(t)\label{eqnr}\\
dD(t)/dt &=& \mu I(t)\label{eqnd}.
\end{eqnarray}
Note that the four equations (\ref{eqns})--(\ref{eqnd}) add up to $0$, reflecting Assumption I(a). The parameters $\beta$, $\nu$ and  $\mu$ which govern the nature of the epidemic are constants, reflecting Assumption I(b). They are known as the \textit{effective contact rate},  \textit{recovery rate} and the  \textit{fatality rate} respectively.
It is clear that the epidemic shall progress if the number of infected
individuals increases over time:
\begin{eqnarray} dI(t)/dt > 0 &\Longleftrightarrow & \beta S(t) I(t)-(\nu+\mu) I(t)> 0\\
 &\Longleftrightarrow & \beta S(t)/(\nu+\mu) > 0\label{propagate}.
\end{eqnarray}
Initially,  everyone, except the first infected case is susceptible. Hence
$S(0)\equiv 1$ So from (\ref{propagate}), for the epidemic to progress we must necessarily have
$$\dfrac{\beta}{\nu+\mu} > 1.$$
On the other hand, it is easy to see that in this SIR(D) model,
$$\tau \tilde c=\beta, \ \ \text{and} \ \  d=(\nu+\mu)^{-1}.$$
This implies that in the SIR(D) model, the reproduction number is nothing but
\begin{equation}\label{eq:rbetanumu}\mathcal{R}_0= \dfrac{\beta}{\nu+\mu}\end{equation}
and thus the epidemic progresses if this quantity is $ >1 $. Therein lies the importance of the reproduction number.

Note that for the SIR(D) model, the reproduction number is a function of the three parameters and does not change over time. For more complicated models, the definition and interpretation of the reproduction number is necessarily much more involved. It may also evolve over time, specially if there are interventions (such as locks and unlocks).


\subsection{Estimation, model fitting and prediction}
As mentioned earlier, we do not try to build a model with components that explicitly accounts for the factors inducing changes to the underlying epidemiological system. Our approach is to start with the standard SIR(D) model and allow it to automatically respond and adjust to the underlying dynamics that arise from such factors.

With respect to the SIR(D) model, note that the data is obtained only at discrete points (in our case daily).
Let
\begin{eqnarray} s(t)&=& \text{increment in susceptible individuals in the}\  t\text{th}\ \ \text{time period},\\
i(t)&=& \text{increment in number of infections in the} \ t\text{th}\ \text{time period},\\
r(t)&=& \text{increment in number of recoveries in the}\ t\text{th}\ \text{time period},\\
d(t)&=& \text{increment in number of fatalities in the} \ t\text{th}\ \text{time period}.
\end{eqnarray}
Moreover, for any choice of the values of $\beta$, $\nu$ and $\mu$, the SIR(D) differential equation can be solved numerically.


How should an SIR(D) be fitted? Many examples of implementation of SIR(D) available in the public domain minimise the residual sum of squares
(RSS) of the  number of infections, to estimate the parameters. This is probably due to the fact that in general,  information is available only on this variable.  However, for COVID-19, information on recoveries and fatalities are also being collected globally. Thus there are other natural candidate RSS. 
At any given time point,
we use an aggregated RSS over number of infections, number of recoveries and number of fatalities. Details are explained below.\vskip5pt

\noindent \textit{Estimate of the parameters $\beta$, $\nu$ and $\mu$}.
As mentioned above we use the RSS. This is done adaptively over a few lagged time points.
Since there are three parameters, one needs data for at least three periods. However, since the data in the initial time periods are sketchy, and also there are smoothness issues, as a general principle, we use a window of past seven days in our estimation procedures. When we are at time point $T$, the initial estimates are obtained by using the least squares on the data for the periods, $T-6, T-5, T-4, T-3, T-2, T-1, T$ {\textemdash}minimizing the sum of squares:
\begin{equation}\label{eq:ss}\sum_{t=T-6}^{T}\big[(i(t)-\hat i(t))^2 +(r(t)-\hat r(t))^2 +(d(t)-\hat d(t))^2\big],
\end{equation}
where $\hat i(t)$, $\hat r(t)$ and $\hat d(t)$, are the observed values of infections, recoveries and fatalities in the  $t$-th period. The values of $i(t)$, $r(t)$ and $d(t)$ for given parameter choices (over a grid) are obtained by numerically solving the differential equations as mentioned above.

The various shocks that the system experiences would also be reflected in the time series of the underlying parameters and hence in their estimates. Thus,
after obtaining the initial estimates, we use a smoother: we chose a weighted average over the last 7 days' estimates with geometrically decreasing weights $0.75^{T-j}$ at time points $j=T-6, \ldots , T$. Other smoothers could also be used. After trial and error, we settled on the value $0.75$.
We thus arrive at our robustified final estimates say, $\beta_T$, $\nu_T$ and $\mu_T$ at the time point $T$.
\vskip5pt

\noindent \textit{Estimate of the reproduction number $\mathcal{R}_0$}:
As mentioned earlier, the \textit{reproduction number}  $\mathcal{R}_0$ is a key quantity that governs the spread of an epidemic and provides a one number summary of the state of the epidemic{\textemdash}even in a situation such as ours where we apply penalization methods and work with a modification of the basic model.
The magnitude of this number indicates whether the epidemic shall spread or die down, and how fast.
Due to the dynamic nature of our model and estimation process, this parameter too is now a time series and requires appropriate estimation procedure.


Recall the relation (\ref{eq:rbetanumu}) between $\mathcal{R}_0$ and the three SIR(D) model parameters. Once estimates of $\beta$, $\nu$ and $\mu$ are obtained as described above, an estimate of $\mathcal{R}_0$  at any time point can be  obtained from (\ref{eq:rbetanumu}) by plugging in the above estimated SIR(D) parameters. However,  this produces a lot of spikes in the estimates. We robustify this estimate by using the median of a moving window of seven past values to obtain our final estimate $\mathcal{R}_{0T}$. This window can be the entire past (conservative) or the recent past (more adaptive but less conservative). We could have used other means of robustification too.

This completes our estimation procedure. Note that the final estimates $\beta_T$, $\nu_T$, $\mu_T$ and $\mathcal{R}_{0T}$ continue to evolve over time. Moreover, at any given time $T$, they do not necessarily satisfy equation (\ref{eq:rbetanumu}). This should not be a cause for concern since after all, the SIR(D) model is not expected to be a perfect model for the epidemic and hence equation (\ref{eq:rbetanumu}) is not sacrosanct.
\vskip5pt

\noindent \textit{Point prediction of $I(\cdot)$, $R(\cdot)$  and $D(\cdot)$}: Given the estimated parameters, the SIR(D) model can then be solved numerically. So, we use our estimates $\beta_{T-1}$, $\mu_{T-1}$ and $\nu_{T-1}$ obtained at time $T-1$ to calculate the $1$-step ahead predicted (solved) values of $I(T)$, $R(T)$  and $D(T)$. We can also calculate the $k$-step ahead predicted values. While solving this, we found it extremely useful to add the obvious but crucial constraint $R(T)+D(T)\leq I(T)$.

\noindent \textit{Prediction bands for $I(\cdot)$, $R(\cdot)$ and $D(\cdot)$}:
Our SIR(D) model, even though dynamic, has no random component.
Thus it does not automatically facilitate calculation of prediction bands. We fall back on a non-parametric method for this purpose. At any given time point $T$, and for any specific variable (infection/recovery/death), we have the past errors (say, from one-step prediction). We can consider the empirical distribution of these $T-1$ error values (or some suitable robust version).
A predictive interval is then obtained by building an interval around the point prediction value using this empirical distribution. It may be noted that if we wish to have a predictive band for the $k$-step ahead predicted value, the width of the band remains same as there is no updated empirical distribution available.

\section{Results}\label{sec:analysis}
\begin{figure}[ht]
\begin{center}
\includegraphics[height=40mm, width=120mm]{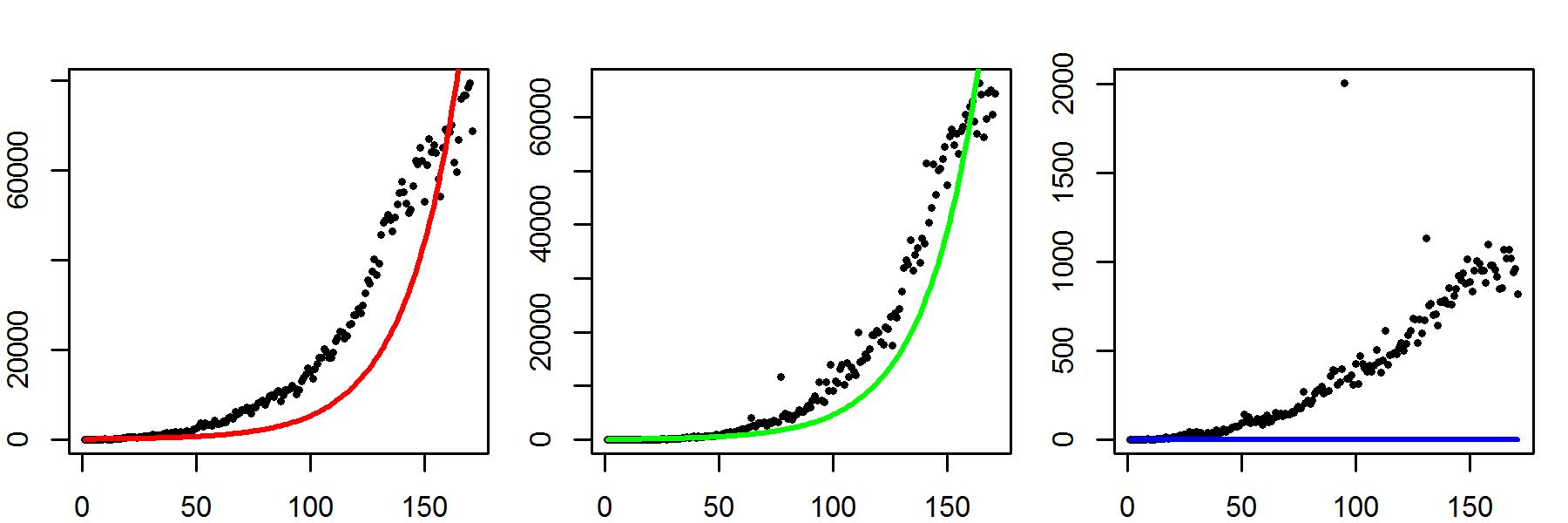}
\small
\caption{\small Aggregated daily counts for India, March 14-August 31, 2020:
infections (left); recoveries (middle); fatalities (right) against number of days. Solid lines:  fitted SIR model.}
\label{fig:India_overall_SIRD_v0.pdf}
\end{center}
\vskip-30pt
\end{figure}

We wish to illustrate our model/method on the COVID-19 data for India.
For the national aggregate level data, multiple data sources are available. Figure \ref{fig:India_overall_SIRD_v0.pdf} exhibits the three time series of number of infections, recoveries and fatalities in India and also illustrates the inadequacy of the traditional unmodified SIR(D) model for this data.

\normalsize

In contrast, with regard to data availability, regional data for finer administrative units (for example the 739 districts of India) is scattered in heterogeneous and/or incompatible formats across numerous local sources. It can be an onerous task to authenticate, clean and collate this data to bring it to a common platform for analysis. Thus, we have used regional data available from \url{https://api.covid19india.org}.
While this data is updated daily, it has idiosyncrasies such as,  missing values, negative counts, delay in updating, change in the reporting format, all of which are expected in a situation as complex as COVID-19. Thus some preliminary consistency checks and cleaning were carried out before proceeding to analyse the data.
\begin{figure}[ht]
\begin{center}\includegraphics[height=100mm, width=125mm]{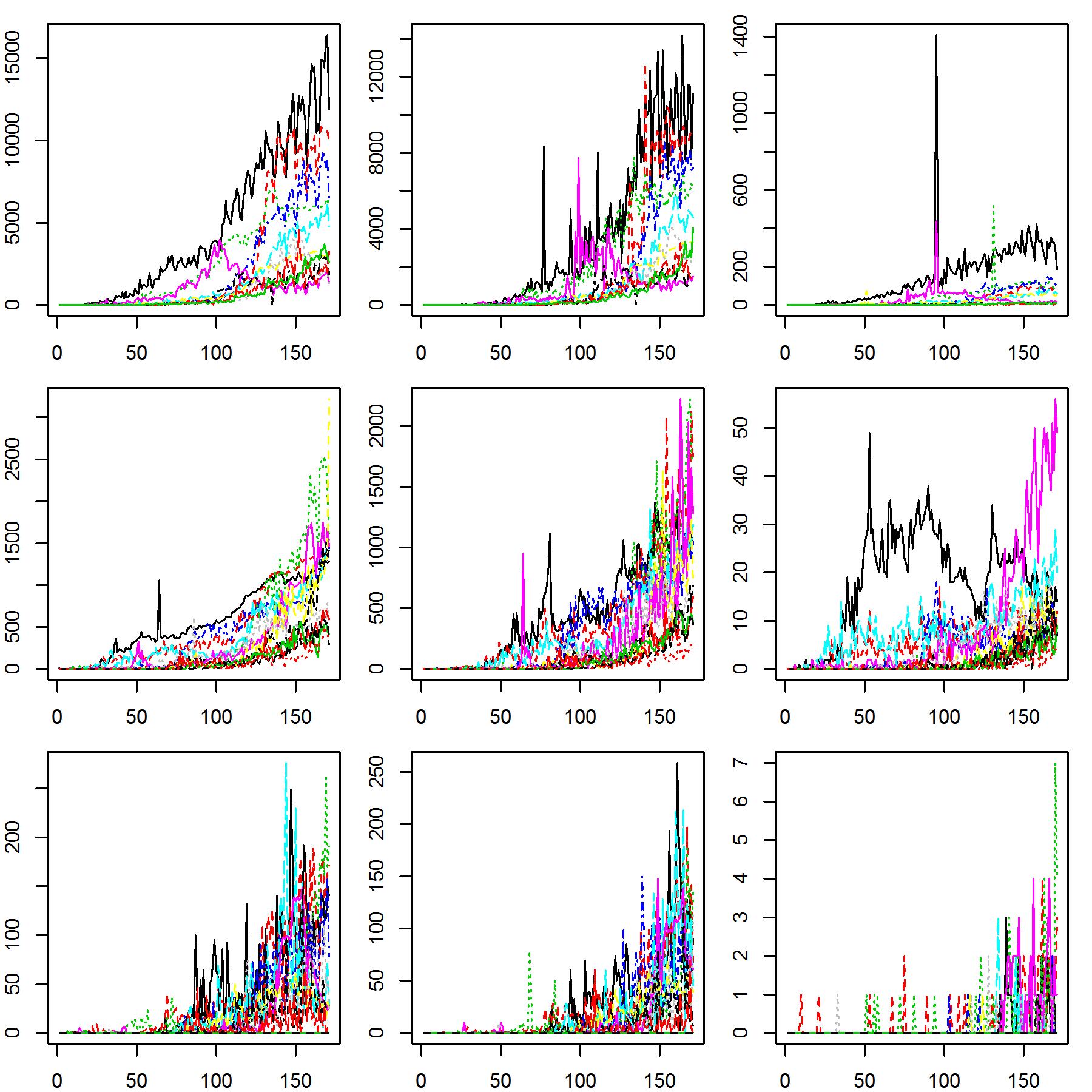}
\caption{\small Daily count of infections (left), recoveries (middle) and fatalities (right) for 36 regions.}
\label{fig:statetimeseriesplots}
\end{center}
\vskip-30pt
\end{figure}
In Figure \ref{fig:statetimeseriesplots} we have plotted the daily count data after dividing the regions into three groups with cumulative infections till August 31, 2020 as: less than 10,000 (12 regions), between 10,000 and 100,000 (13 regions) and above 100,000 (11 regions).

\begin{figure}[ht]
\vskip-10pt
\begin{center}
\includegraphics[height=50mm, width=150mm]{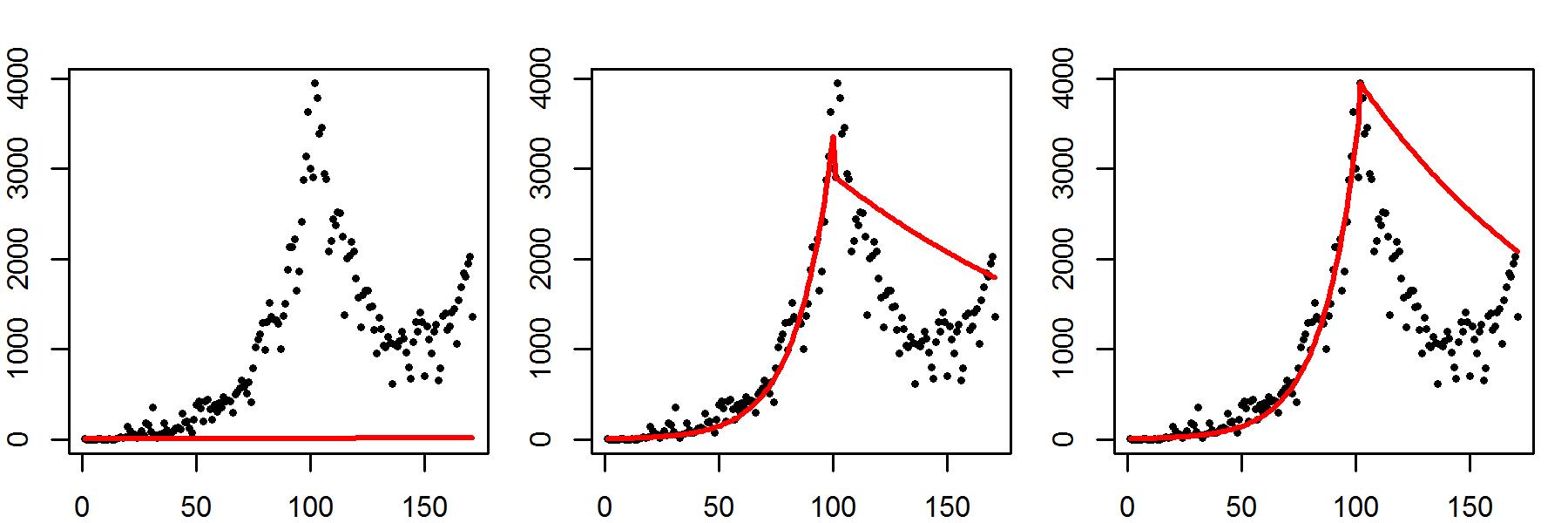}
\end{center}
\vskip-10pt
\caption{\small Performance of different  SIR(D) on Delhi data.
Left: single SIR(D). Middle and right: SIR(D) with change points at 100-th day and 101-th day respectively.}
\label{fig:India_state_SIRD_Delhi_v0.pdf}
\end{figure}

As mentioned earlier, due to known and hidden factors, a simple SIR(D) model cannot be a good fit and one needs to be adaptive to the changes in the spread. Figure \ref{fig:India_state_SIRD_Delhi_v0.pdf} illustrates this on Delhi data. The solid line in the left panel corresponds to the fitted values obtained using a single SIR(D) and overlayed with the observed data clearly shows model inadequacy. The middle and right panels use a modified SIR(D) to accommodate a single (known) change point{\textemdash}at the 100th day (middle panel) and at the 101th day (right panel). While the fit improves somewhat, vulnerability to the choice of change point is evident and moreover, the fit is still unsatisfactory.
\vskip-15pt
\begin{figure}[ht]
\begin{center}\includegraphics[height=80mm, width=100mm]{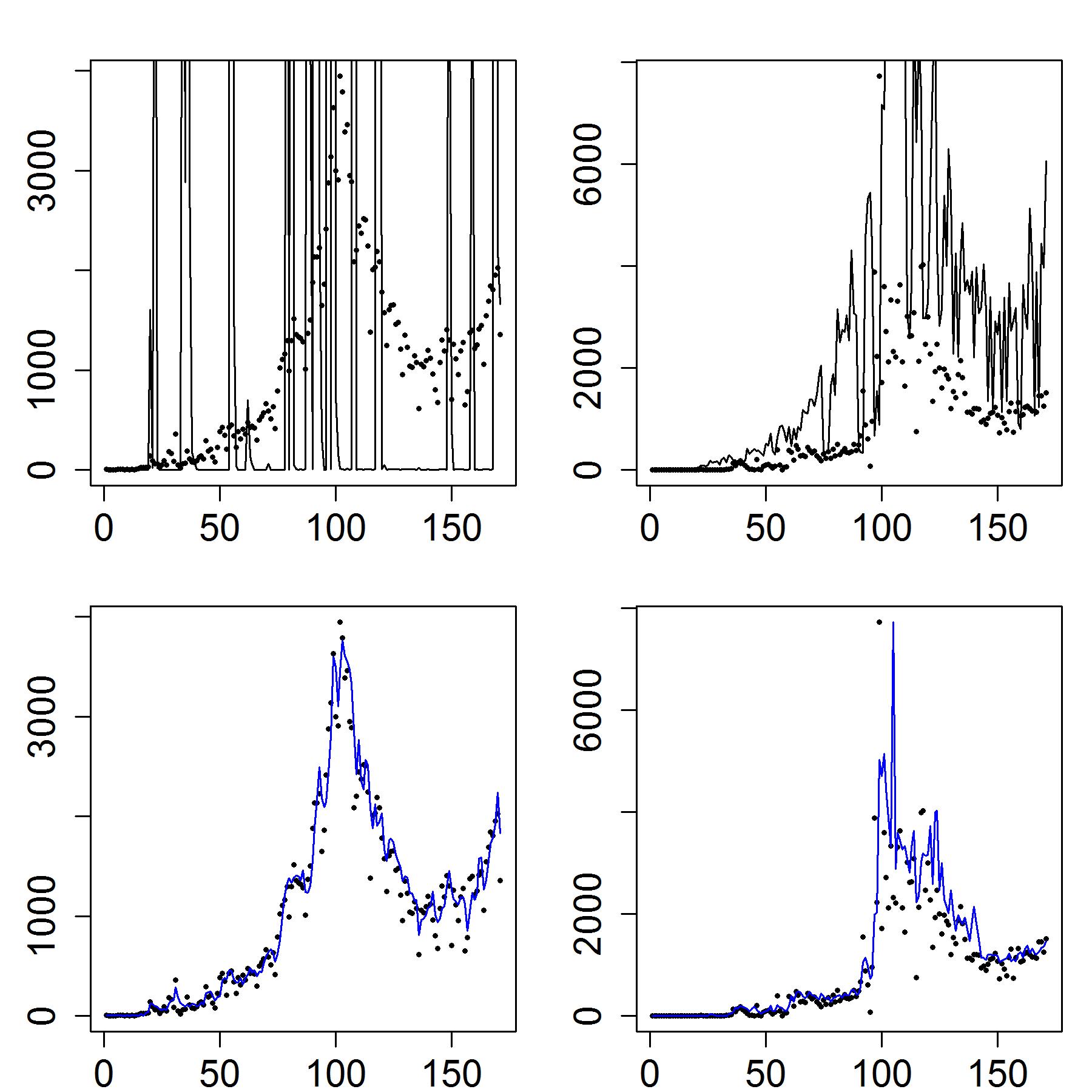}
\end{center}\vskip-20pt\caption{\small Fit with different RSS criteria. Top panel: partial RSS. Bottom panel: total RSS. Left and right panels: number of daily infections and recoveries respectively.}
\label{fig:India_state_SIRD_Delhi_v1.pdf}
\vskip-20pt
\end{figure}

As already pointed out,  making the model dynamically adaptive alone may not be sufficient. Figure \ref{fig:India_state_SIRD_Delhi_v1.pdf} on the Delhi data illustrates the improved performance as we incorporate additional penalties for lack of model fit. The plots in the bottom panel, obtained by our proposed method, are in sharp contrast to those in the top panels which were obtained by commonly used partial penalty methods. These plots clearly demonstrate the advantage of optimising over the combined RSS with respect to all three variables, over a window of data, yielding a dynamic SIR(D) model. The observed and the fitted values are seen to be in excellent agreement while this dynamic SIR(D) continues to be parametric.

Since the proposed SIR(D) model is allowed to evolve over time, estimate of these parameters are updated daily by refitting the model taking into account the additional observations for the day.  \vskip5pt

\begin{figure}[ht]
\begin{center}\includegraphics[height=100mm, width=100mm]{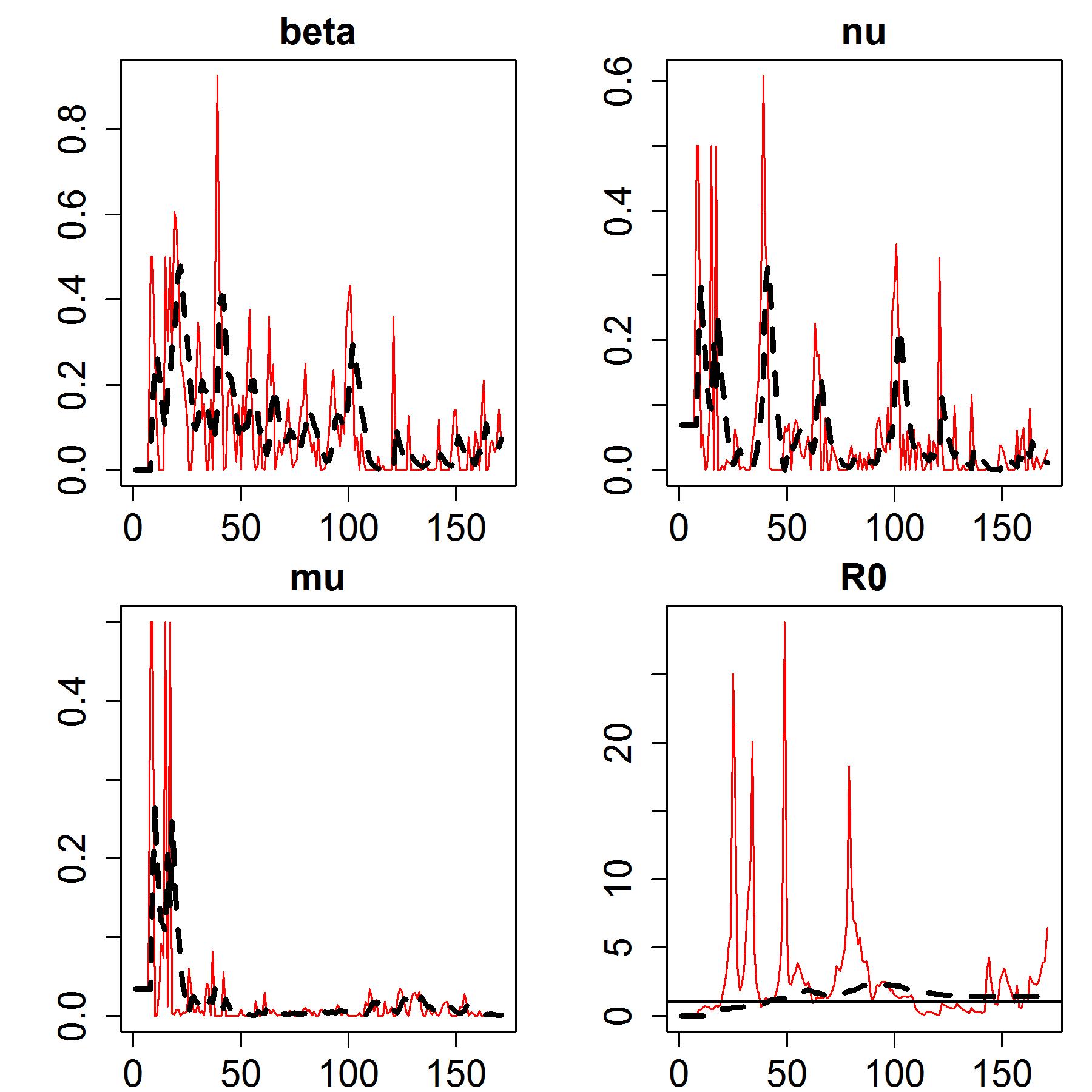}
\caption{\small Initial (in red) and final (in black) estimates of $\beta$, $\nu$, $\mu$ and  $\mathcal{R}_0$.}
\label{fig:India_state_SIRD_Delhi_v2.pdf}.
\end{center}
\vskip-20pt
\end{figure}

As explained in Section \ref{sec:background}, there are several local and  general shocks and perturbations that the underlying epidemiological system experiences. These are naturally reflected in the time series of the underlying parameters. Our dynamic model captures these and we notice some changes in the parameter estimates as well. Thus, after obtaining the initial estimates, we use a weighted average over the estimates from the past week with decreasing weights. As an illustration, the spiky and smoothed estimates of the three parameters and the robust estimate of $\mathcal{R}_0$ for Delhi are  plotted in Figure
\ref{fig:India_state_SIRD_Delhi_v2.pdf}.

\begin{figure}[ht]
\begin{center}
\includegraphics[height=100mm, width=150mm]{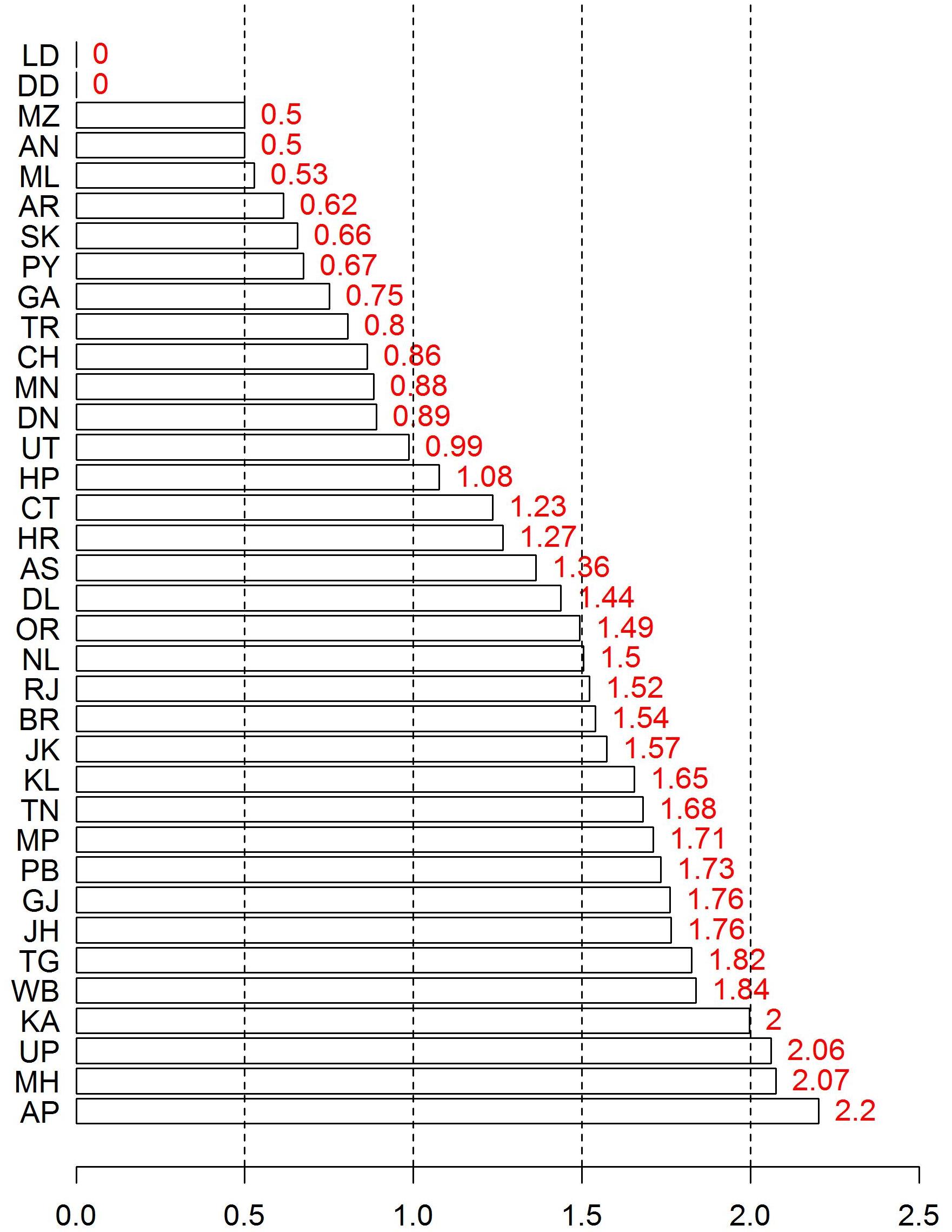}
\caption{\small Region-wise estimate of the reproduction number as on August 31, 2020.}
\label{fig:India_state_level_current_R0.pdf}
\end{center}\vskip-25pt
\end{figure}

We implemented the above in the software {\tt R}. For solving the primary differential equations of the SIR(D) model numerically,  we used the standard library, {\tt desolve}  within {\tt R}. Additional codes required to carry out robust estimation and prediction were also implemented in {\tt R}. The entire programme takes a few minutes to run on an Intel core i3, 2.4GHz computer with past 6 months data on all 36 Indian regions. Detailed plots and figures are given in the  Supplementary file. We present briefly the highlights:

(i) \textbf{Reproduction number}: Figure \ref{fig:India_state_level_current_R0.pdf} gives the estimate for the current reproduction number, as on 31-Aug-2020,  for all the regions.

Table \ref{table:r0freq} gives a summary frequency distribution. Three regions, Maharashtra, Andhra Pradesh, and Uttar Pradesh are currently the worst placed.

\begin{table}[h]
\begin{center}
\caption{\small Summary distribution of reproduction number as on August 31, 2020}\label{table:r0freq}
\vskip 3pt
{\footnotesize
\begin{tabular}{|c|c|}
\hline
Estimated $\mathcal{R}_0$     &  Number of regions             \\
 \hline
             Effectively zero &  02         \\
 $ 0 < \mathcal{R}_0\leq 0.5$ &  02         \\
$0.5 < \mathcal{R}_0 \leq 1$  &  10           \\
$1 < \mathcal{R}_0 \leq 1.5$  &  06           \\
$1.5 < \mathcal{R}_0 \leq 2$  &  13           \\
$ \mathcal{R}_0 > 2$          &  03           \\
\hline
Total                         &  36           \\
\hline
\end{tabular}}
\end{center}\vskip-15pt
\end{table}

\begin{figure}[ht]
\begin{center}
\includegraphics[height=110mm, width=120mm]{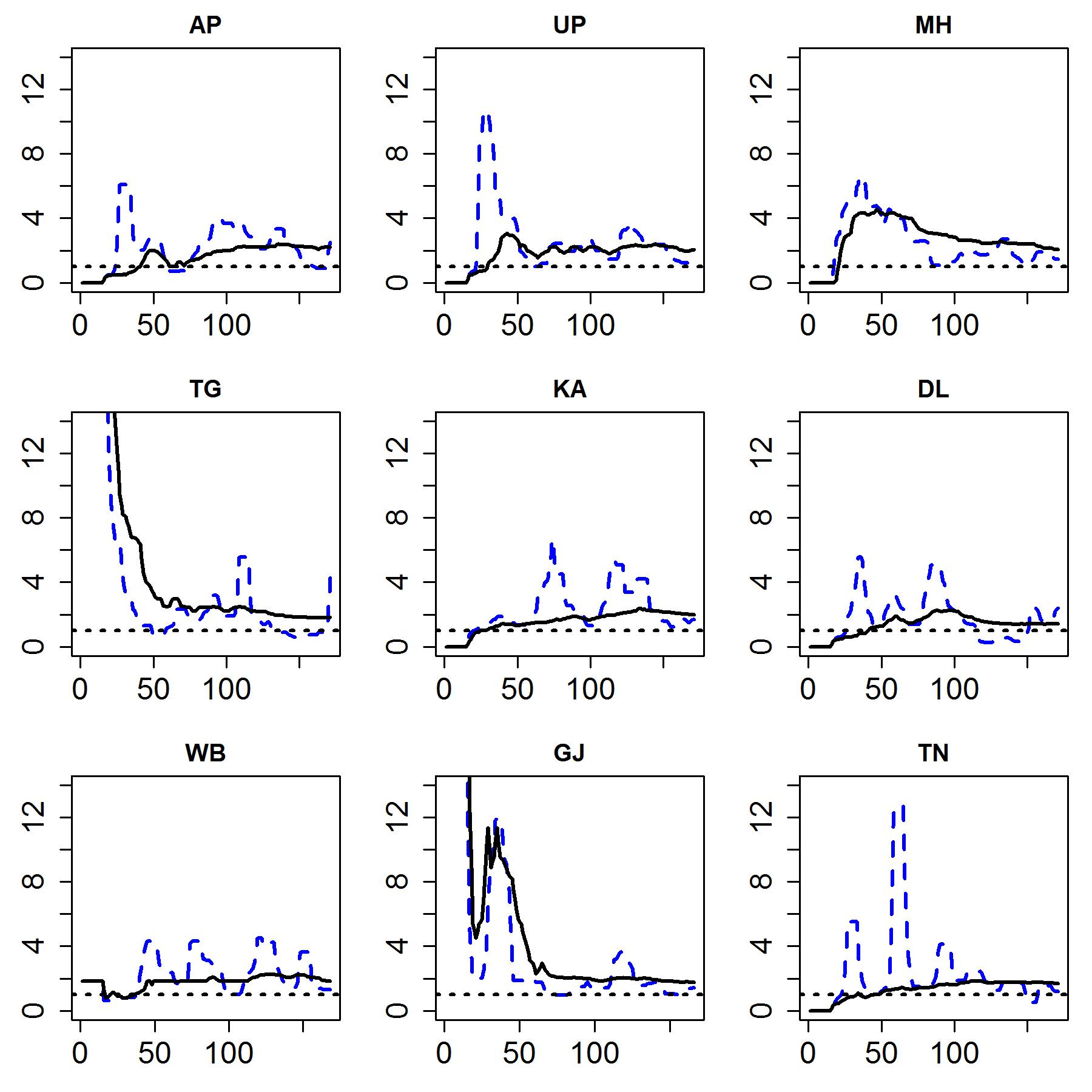}
\caption{\small Estimated time series of $\mathcal{R}_0$ with  moderate and conservative methods for some regions with highest infections.}
\label{fig:Figure_R0_sel9_states.jpg}
\end{center}
\vskip-25pt
\end{figure}
Plots for the time varying  estimates of $\mathcal{R}_{0}$ for nine regions with highest number of infections are given in Figure \ref{fig:Figure_R0_sel9_states.jpg}. Plots for all the regions can be found in  the Supplementary file.  We have robustified our basic estimate by considering the median over the previous two weeks. We also tried a more conservative version by taking the median over the entire past. We have already noted that most regions seem to have evolved in their individual ways and this observation continues with respect to $\mathcal{R}_{0}$. However there are similarities. For instance, there is typically an initial phase of shock(s) followed by a relatively steady (and lower) level of $\mathcal{R}_0$. For some regions it appears that there is a gradual decreasing trend as well, indicating that the peak of the pandemic may be in sight over the coming months in these regions.

\begin{figure}[ht]
\begin{center}
\includegraphics[height=75mm, width=120mm]{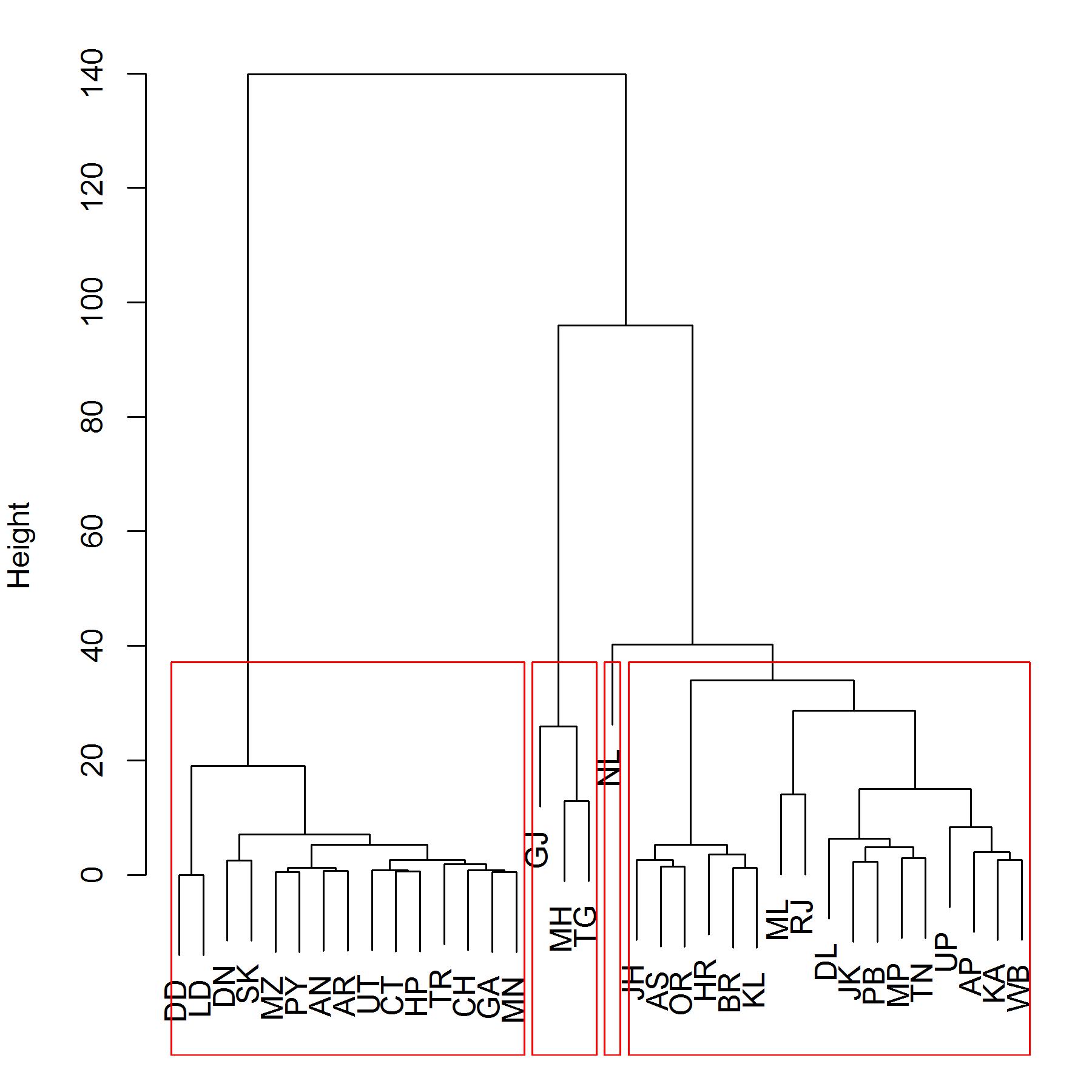}
\caption{\small Dendrogram showing clusters based on time series of reproduction number.}
\label{fig:India_state_level_R0_TScluster.pdf}
\end{center}\vskip-25pt\end{figure}

As we mentioned above, the regions have evolved differently over time and this becomes evident when we carry out a time series clustering.
We use the method of Ward (1963) which minimises the between-cluster Euclidean distance at each sequential step.
The height of the fusion given on the $Y$-axis, indicates the dissimilarity between two observations.
The proximity of two observations along the horizontal axis does not signify any closeness.
Figure \ref{fig:India_state_level_R0_TScluster.pdf} shows the clusters of regions based on the time series of reproduction numbers.

(ii) \textbf{Prediction}: We have calculated the daily predicted values of the infected, recovered and deceased, along with 99\% predictive intervals.
Note that the derivation of the predictive interval requires estimation of the error distribution. This has been done non-parametrically by using the past errors in prediction. We illustrate this
with the Delhi data since this region shows sharp turns in the daily values and thus is relatively difficult to predict. Prediction for all the regions is available in the Supplementary file, along with the estimated predictive error distribution in each case.
\begin{figure}[ht]
\begin{center}
\includegraphics[height=50mm, width=120mm]{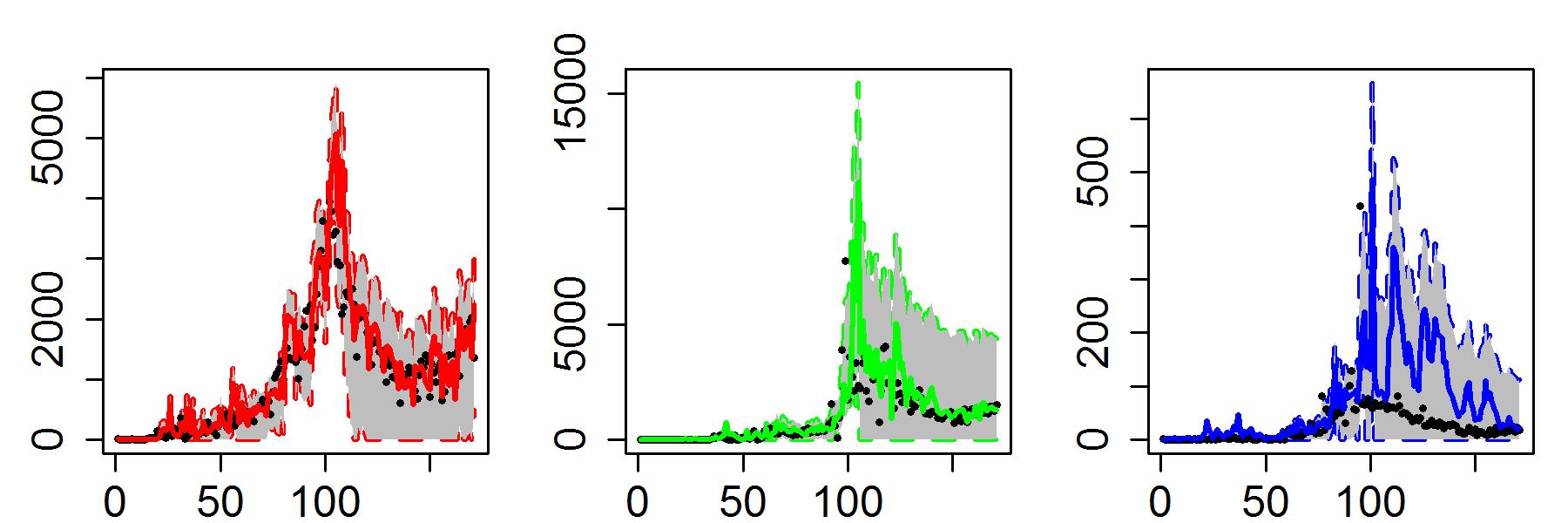}
\caption{\small Predictive band of daily counts for Delhi: infections (left), recoveries (middle) and fatalities (right).}
\label{fig:delhipredictbands}
\end{center}
\vskip-25pt
\end{figure}

\begin{figure}[ht]
\begin{center}
\includegraphics[height=100mm, width=100mm]{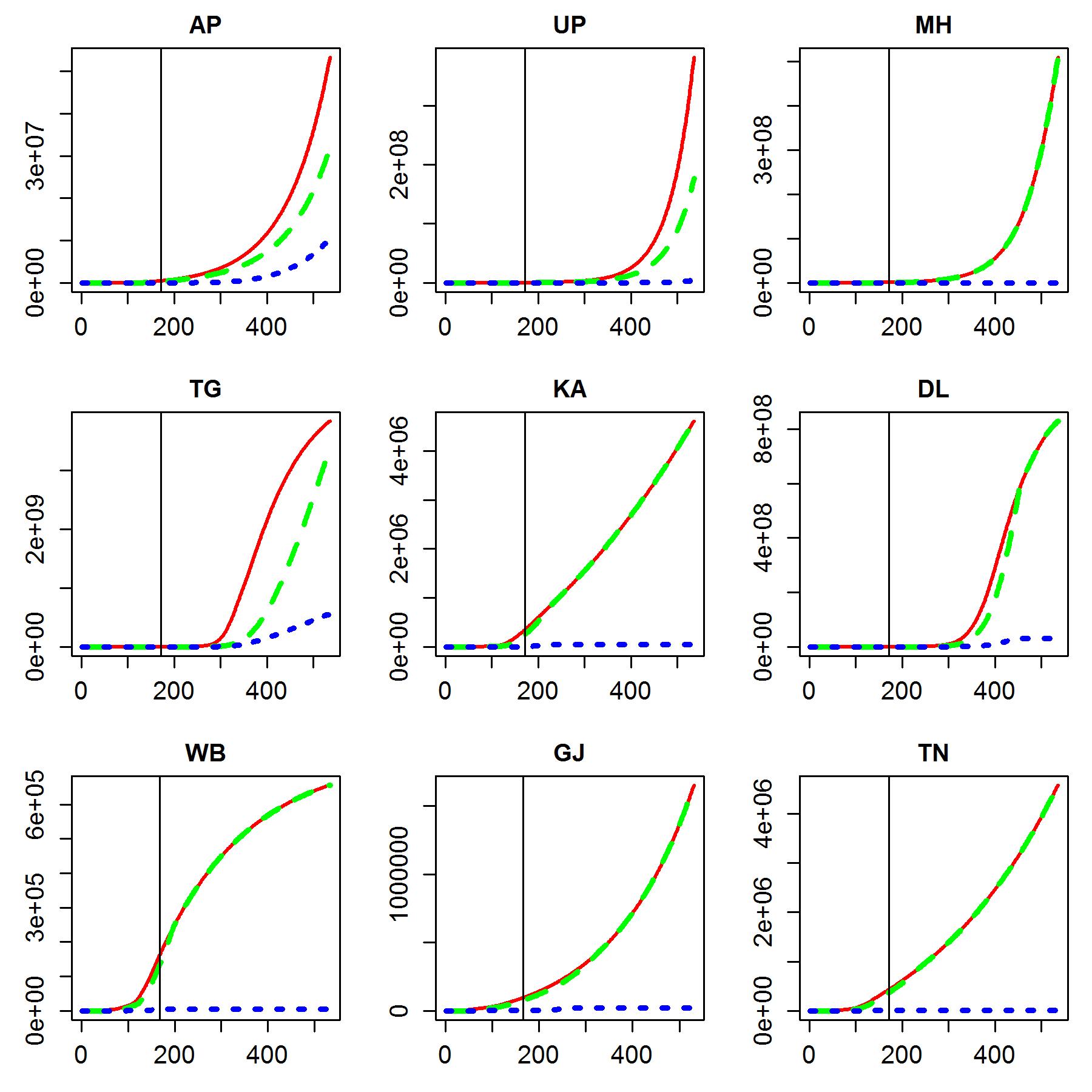}
\caption{\small Long term predicted counts as of August 31, 2020, of infections (solid lines), recoveries (long-dashed lines) and fatalities (short-dashed lines) for some regions.}
\label{fig:longterm}
\end{center}
\vskip-25pt
\end{figure}

We have also carried out one full year's prediction to assess the time and extent of future peaks.
Figure \ref{fig:longterm} gives the plots for nine regions. These regions were chosen since they have the maximum number of infections as of August 31. See the Supplementary file for values for all regions.

\noindent
It may be noted that in several regions, the number of infections and fatalities are predicted to rise significantly. The heterogeneity in these figures across  regions is also very noticeable. In several regions, the pandemic is far from over and have not seen their peaks. However, there is reason for optimism since in some regions the spread seems to be slowing down.  \vskip5pt


\section{Discussion}\label{sec:conclusions}
We have considered a dynamic version of the time tested SIR(D) model and fitted it to the Indian regional COVID-19 data. This has been done by minimising aggregated RSS dynamically over sliding windows.  Appropriate smoothers and robustification methods have been used to obtain time series estimates of the \textit{effective contact rate}, the \textit{recovery rate}, the  \textit{fatality rate} and the \textit{reproduction number}. We have illustrated our method on the COVID-19 Indian regional data. Out of the 36 regions, 20 have a reproduction number higher than 1.0 and three of these are higher than 2.0.  We also use the fitted model for short and long term prediction of the number of infection, recoveries and fatalities. The projected pattens are quite varying by regions and while the prognosis is good for some regions in others a sharp rise is foreseen.

Progression of COVID-19 in India is impacted by spatial dependence and it would be pertinent to develop a spatio-temporal model with regions or even districts as the primary units. We plan to investigate this by developing a suitable spatio-temporal model. Studying the effects of migration is a particularly challenging task.

\vskip10pt

\noindent \textbf{References}\vskip3pt

\textit{Data source}: https://api.covid19india.org (also at https://github.com/covid19india)

\textit{Supplementary file} (supplementary.pdf): Six plots for each of the 34 regions with non-trivial counts:

\hskip10pt (a) daily infections: observed value and one-step prediction with 99\% prediction band.

\hskip10pt
(b) daily infections: predictive error distribution.

\hskip10pt
(c) daily recoveries: observed value and one-step prediction with  99\% prediction band.

\hskip10pt
(d) reproduction number estimates: smoothed over past 14 days, and the entire past.

\hskip10pt
(e) daily fatalities: observed value and one-step prediction with 99\% prediction band.

\hskip10pt
(f) cumulative infected, recovered, fatalities:  predicted values till one year, beginning \\
\text{}\hspace{24pt} September 01, 2020.
\vskip10pt

Ansumali, Santosh and Prakash, Meher K. (2020). A very flat peak:
why standard SEIR models miss the plateau of COVID-19 infections and
how it can be corrected. \textit{MedarXiv}. \url{https://doi.org/10.1101/2020.04.07.20055772}

Clif, A. D.  and Ord, J. K. (1981). \textit{Spatial processes: models and applications}. Pion.

ET (2020). Over 11.2 lakh Indians have returned from abroad under Vande Bharat Mission: MEA. News item, August 20, 2020. \textit{Economic Times}. \url{https://economictimes.indiatimes.com/news/politics-and-nation/over-11-2-lakh-indians-have-}\\ \url{returned-from-abroad-under-vande-bharat-mission-mea/articleshow/77661541.cms}. [Accessed August 25, 2020.]

Ghosh, Abantika and Kapoor, Shipra (2020). 45,720 new cases in 24 hrs, 13\% positivity, single-day toll 1,129 after Tamil Nadu update. News item.  \textit{The Print}.
\url{https://theprint.in/health/45720-new-cases-in-24-hrs-13-positivity-single-day-}\\ \url{toll-1129-after-tamil-nadu-update/466882/} [Accessed August 17, 2020.]

HT (2020a). India’s death toll soars past 10K, backlog deaths raise count by 437 in Delhi, 1,409 in Maharashtra. News item. \textit{Hindustan Times}. \url{https://www.hindustantimes.com/india-news/india-s-death-toll-soars-past-10k-backlog-deaths-raise-}\\ \url{count-by-437-in-delhi-1-409-in-maharashtra/story-9GNbe7iMBKLsiHtByjRKCJ.html}. [Accessed August 17, 2020.]

HT (2020b). India extends ban on international flights till August 31. News item, July 31, 2020. \textit{Hindustan Times}. \url{https://www.hindustantimes.com/india-news/india-}\\ \url{extends-ban-on-international-flights-till-august-31/story-} \\ \url{ervIbqmVguo9xNEQxozWJL.html}.  [Accessed August 25, 2020.]

Kermack, W.O., McKendrick, A.G. (1991a). Contributions to the mathematical theory of epidemics--I. \textit{Bulletin of Mathematical Biology} 53, 33--55 (1991). \url{https://doi.org/10.1007/BF02464423}.

Kermack, W.O., McKendrick, A.G. (1991b). Contributions to the mathematical theory of epidemics--II. The problem of endemicity. \textit{Bulletin of Mathematical Biology} 53, 57--87. \url{https://doi.org/10.1007/BF02464424}.

Kermack, W.O., McKendrick, A.G. (1991c). Contributions to the mathematical theory of epidemics--III. Further studies of the problem of endemicity. \textit{Bulletin of Mathematical Biology} 53, 89--118. \url{https://doi.org/10.1007/BF02464425}.

Kotwal,  Atul; Yadav, Arun Kumar; Yadav, Jyoti; Kotwal,  Jyoti and Khune, Sudhir (2020). Predictive models of COVID-19 in India: A rapid
review.  \textit{Medical Journal Armed Forces India}. To appear in print. Available online June 17, 2020. \url{https://doi.org/10.1016/j.mjafi.2020.06.001}.

Kucharski, A. J., et al.
(2020).
Early dynamics of transmission and control of COVID-19: a mathematical modelling study.
\textit{The Lancet Infectious Diseases}, \url{doi: 10.1016/S1473-3099(20)30144-4}.

Liu, Y. , Gayle, A. A., Wilder-Smith,   A.,  and Rocklov, J.  (2020). The reproductive number of COVID-19 is higher compared to SARS coronavirus. \textit{Journal of Medical Virology}, 27(2),
\url{doi:10.1093/jtm/taaa021}.

Mohan, Pavitra  and  Amin, Arpita (2020). Forcing migrants to stay back in cities during lockdown worsened spread of coronavirus, study shows. News item, July 03, 2020. \textit{Scroll.in}. \url{https://scroll.in/article/966123/forcing-migrants-to-stay-back-}\\ \url{in-cities-during-lockdown-worsened-spread-of-coronavirus-study-shows}. \\ {}[Accessed August 25, 2020.]

Ranjan, Rajesh  (2020). COVID-19 spread in India: dynamics, modeling, and future projections. \textit{Journal of the Indian Statistical Association}, Vol.58 No.1, 47-65.

Sun, K., Chen,  J. and Viboud, C. (2020). Early epidemiological analysis of the coronavirus disease 2019 outbreak based on crowd sourced data: a population-level observational study. \textit{The Lancet Digital Health}, \url{doi: 10.1016/S2589-7500(20)30026-1}.

Ward, J. H., Jr. (1963). Hierarchical Grouping to Optimize an Objective Function. \textit{Journal of the American Statistical Association}, 58, 236-244.


The Wire (2020). Here's why phase 4 of lockdown has seen a sudden surge in cases in the Northeast. News item.
\url{https://thewire.in/government/covid-19-case-surge-}\\ \url{northeast COVID-19}.  [Accessed August 25, 2020.]

Wikipedia (2020). Timeline of COVID-19 in India (2020). \url{https://en.wikipedia.org/wiki/Timeline_of_the_COVID-19_pandemic_in_India}. [Accessed August 25, 2020.]

%

\small

\end{document}


\begin{center} Supplement to\\
{\sc Modelling COVID-19 - I\\
A dynamic SIR(D) with application to Indian data}\\
by \\
Madhuchhanda Bhattacharjee \ and \ Arup Bose
\end{center}

This file contains the analysis for 34 regions that have non-trivial infection counts as on August 31, 2020. Abbreviated name of the region along with date of first confirmed case in that region is reported in top left corner. For most figures the X-axis represents number of days since the first confirmed case in that region (other than the figure on top right).

In particular, for each state, we have  six plots as follows: \\

\hskip10pt 
Left panel, top: Daily infections: observed value and one-step prediction with 99\% prediction band.\\

\hskip10pt
Left panel, middle: Daily recoveries: observed value and one-step prediction with  99\% prediction band.\\

\hskip10pt
Left panel, bottom: Daily fatalities: observed value and one-step prediction with 99\% prediction band.\\

\hskip10pt
Right panel, top: Daily infections: predictive error distribution. \\

\hskip10pt
Right panel, middle: Reproduction number estimates: smoothed over past 14 days, and the entire past. \\

\hskip10pt
Right panel, bottom: Cumulative infected, recover September 01, 2020 (marked by a vertical line).

\newpage
\includepdf[pages={1-34}]{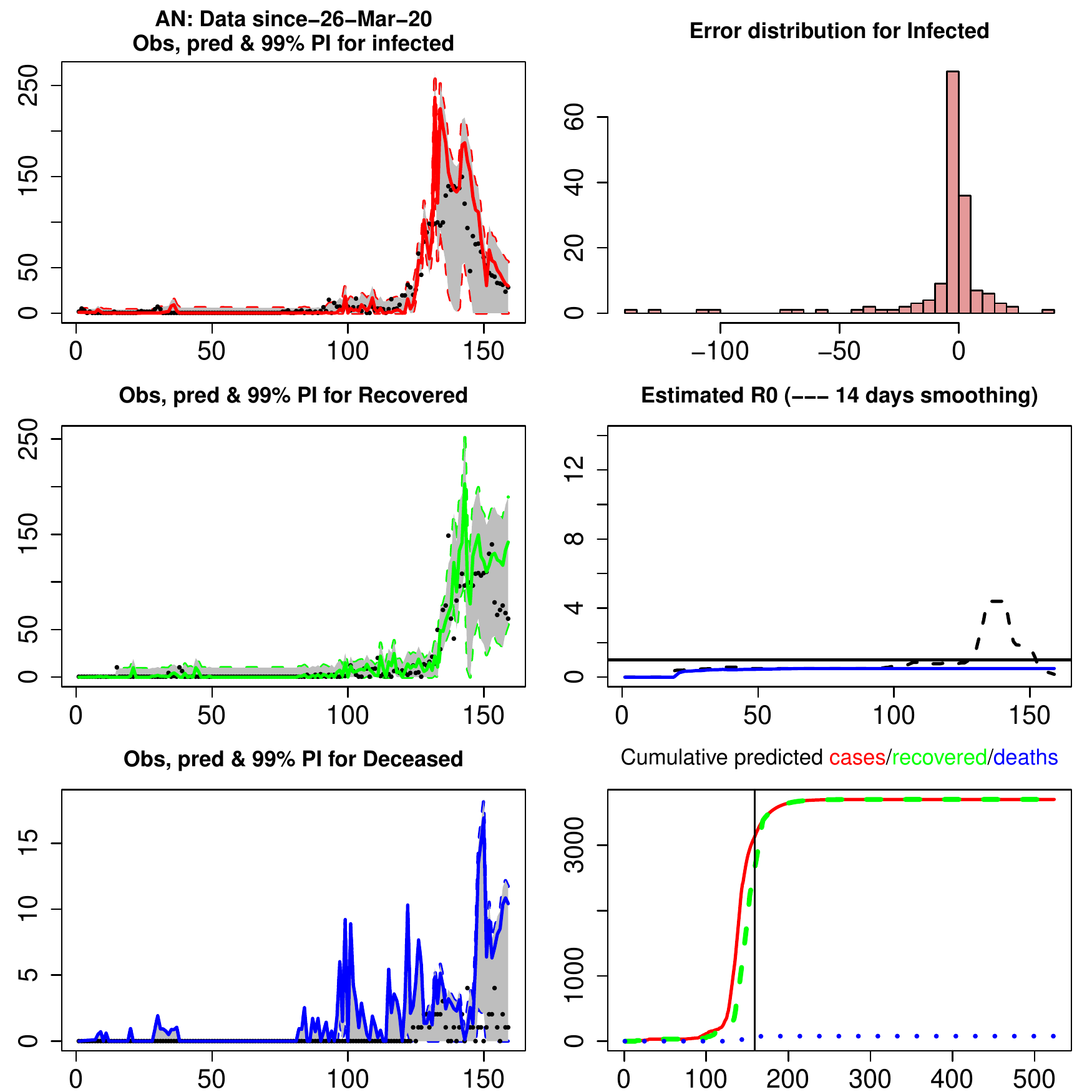}

\vskip10pt